# Influence of interfacial roughness amplitude and periodicity on giant chiral asymmetry of domain wall mobility in epitaxial Cu/Pd/Co/Pd(111) films

*N. N. Chernousov*[1], *A. S. Pashenko*[1], *A. G. Kozlov*[1], *A. A. Turpak*[1], *A. S. Samardak*[1], *and A.V. Davydenko*[1,*]

[1]Laboratory of Spin-Orbitronics, Institute of High Technologies and Advanced Materials, Far Eastern Federal University, Vladivostok, 690922, Russia

Keywords: chiral pinning; chiral damping; Dzyaloshinskii–Moriya interaction; interfacial roughness; domain-wall dynamics; perpendicular magnetic anisotropy

*davydenko.av@dvfu.ru, 0000-0002-2697-8563

## Abstract

We investigate the influence of interfacial roughness amplitude and lateral periodicity on the chirality-dependent mobility of domain walls (DWs) in epitaxial Si(111)/Cu(2–7 nm)/Pd(3 nm)/Co(0.7 nm)/Pd(3 nm) heterostructures. By varying the Cu buffer layer thickness, we systematically modify the morphology of the magnetic interfaces while maintaining a constant composition and thickness of the Pd/Co/Pd trilayer. Scanning tunneling microscopy reveals that increasing the Cu thickness leads to the development of a hierarchical interfacial morphology characterized by increasing roughness amplitude and lateral periodicity. We observe a dramatic enhancement of the asymmetry between the velocities of DWs with opposite chirality in the creep regime, reaching a velocity ratio of approximately 50000 in the sample with a 7-nm-thick Cu buffer layer. This value exceeds by more than an order of magnitude the effective saturated asymmetry previously observed in Pd/Co/Pd structures with roughness controlled by the bottom Pd layer thickness. Despite comparable root-mean-square interface roughness in both series, the Cu-buffer-induced morphology exhibits a substantially larger characteristic lateral periodicity. We propose that the enhanced asymmetry originates from the different spatial organization of interfacial disorder, which modifies the pinning energy landscape and amplifies chirality-dependent thermally activated DW propagation. Our results suggest that engineering not only the amplitude but also the lateral spatial correlations of interfacial roughness provides an effective approach for controlling chiral DW dynamics in magnetic heterostructures.

## 1 Introduction

In the era of artificial intelligence and neuromorphic computing [1], thin magnetic films with perpendicular magnetic anisotropy (PMA), in which DWs can be controllably displaced by external stimuli, are of considerable interest [2]. Such magnetic systems offer a platform for both racetrack memory devices [3, 4] and neuromorphic computing architectures, in which DWs or magnetic domains act as quasiparticles moving along magnetic tracks, analogous to the transport of electrical signals in biological neurons [5]. By merging or branching magnetic tracks, elementary logic units can be constructed and subsequently integrated into more complex neuromorphic architectures [6].

Controlled displacement of magnetic textures in systems with PMA commonly relies on a combination of two effects: the Dzyaloshinskii–Moriya interaction (DMI) [7, 8] and spin–orbit torque (SOT) [9, 10]. When the DMI is sufficiently strong, the DWs acquire a homochiral Néel configuration, with the sense of magnetization rotation determined by the sign of the effective DMI [2]. Under the action of SOT, homochiral DWs of both the $\downarrow\uparrow$ and $\uparrow\downarrow$ types move in the same direction, a property that can be exploited in racetrack logic devices [6].

One direct manifestation of the DMI in thin films with PMA is the difference in DW mobilities under simultaneously applied out-of-plane (OOP) and in-plane (IP) magnetic fields: the OOP field drives DW motion, whereas the IP field breaks the symmetry of the DW configurations [11, 12, 13].

More recently, another mechanism was identified that can produce asymmetric magnetic-domain expansion under combined OOP and IP fields even in systems with negligible DMI [14]. This effect was also demonstrated for DW displacement driven by SOT [15]. Racetrack devices exploiting the large difference in DW mobilities have been proposed to enable accurate and controllable positioning of DWs [14]. The chirality-dependent mobility in systems without DMI has been attributed to chiral damping [16]. Evidence for chiral damping in Cu/CoFeB/W was obtained by Brillouin light scattering (BLS) measurements [17]. The chirality dependence of DW mobility may arise from both DMI and chiral damping, and separating their contributions remains challenging [18]. Recently, we observed a giant mobility difference between DWs of opposite chirality in epitaxial Pd/Co/Pd(111) trilayers [19]. The velocity ratio between left- and right-handed Néel DWs reached several thousand, hundreds of times larger than the relative velocity differences typically associated with DMI alone. The mobility asymmetry was found to increase monotonically with the bottom Pd layer thickness. At the same time, the roughness of the Pd/Co and Co/Pd interfaces and the coercive field of the Co layer increased substantially. These findings indicated an important role of the interaction between DWs and structural disorder in Pd/Co/Pd. In a subsequent study, we demonstrated that the chirality dependence of the mobility is most pronounced in the creep regime, where DW motion is governed by pinning at structural defects, and is strongly suppressed upon approaching the depinning regime [20].

Here, we investigate Si(111)/Cu(2–7)/Pd(3)/Co(0.7)/Pd(3) heterostructures (all layer thicknesses are given in nanometers unless stated otherwise). Our initial objective was to determine whether giant DW mobility asymmetry could be achieved by controlling interfacial roughness through the thickness of the Cu buffer layer, while keeping the bottom Pd layer thickness fixed. We find that this approach provides substantially more efficient control over DW mobility asymmetry than varying the bottom Pd thickness. Surprisingly, despite comparable root-mean-square (rms) roughness values, the two series exhibit markedly different degrees of mobility asymmetry. A comparative analysis of the interface morphology reveals a substantial difference in the characteristic lateral periodicity of roughness, arising from different film growth mechanisms. These findings suggest that not only the amplitude but also the lateral spatial organization of interfacial roughness plays an important role in chirality-dependent DW pinning and propagation.

### 1.1. Experimental Details

Si(111) substrates were cut into 12 × 4 $mm^2$ pieces and cleaned in acetone, isopropanol, and deionized water before being introduced into the vacuum chamber. The substrates were then degassed in situ at 700–800 °C for 12 h. Immediately before deposition, the native oxide was removed by applying a short heating-current pulse of 11 A. Cu, Co, and Pd were deposited from effusion cells at rates of 0.45, 0.28, and 0.26 nm/min, respectively. The deposition rates were monitored using a quartz-crystal microbalance calibrated by reflection high-energy electron diffraction (RHEED). The substrate temperature was 340 K during deposition of Cu and the bottom Pd layer and 370 K during deposition of Co and the top Pd layer. These temperature differences originated from differences in infrared heating by the effusion cells. The crystalline structure of the growing layers was characterized in situ by RHEED, and their surface morphology was examined by scanning tunneling microscopy (STM). Magnetic hysteresis loops were measured using a vibrating sample magnetometer (Lakeshore) with applied fields up to 2.2 T. Magnetic-domain structures were imaged using a magneto-optical Kerr-effect (MOKE) microscope (Evico-Magnetics) equipped with a custom-built large OOP coil, an OOP microcoil, and an IP electromagnet. The large OOP coil generated field pulses with amplitudes up to 60 mT and a fixed duration of 2 ms for DW velocity measurements in the creep regime. The microcoil generated driving OOP fields with amplitudes up to 150 mT and was used to investigate DW motion during the transition from creep to depinning. The microcoil had a diameter of 2 mm and consisted of two turns of Cu wire. It was mounted on a textolite plate with a central aperture allowing polarized light to illuminate the sample and the reflected light to reach the CCD camera. The microcoil pulses were not rectangular; consequently, the driving field and instantaneous DW velocity varied during each pulse. Simply dividing the DW displacement by the pulse duration would therefore yield only an average velocity. We instead employed a mathematical correction procedure to determine the instantaneous DW velocity at a specified magnetic field. This procedure involves calculating the function $\mathbf{T}(H, v)$, which describes the time density associated with the instantaneous field values during a pulse. The microcoil drive circuit, calibration, and velocity-correction procedure are described in a separate study [21].

Artificial nucleation centers with a diameter of 10 μm were fabricated by focused ion-beam (FIB) etching. For the DW velocity measurements, the sample was positioned so that a nucleation center lay at the center of the large coil or microcoil. After saturating the sample in a negative OOP field, a circular reversed domain was nucleated around this center. A constant IP magnetic field was then applied, followed by a sequence of positive OOP field pulses that displaced the DWs by a distance smaller than the radius of the microscope field of view. The displacement of the DW along the IP field direction was determined from differential-contrast MOKE images.

## 2. Results

### 2.1 Structural Characterization

We investigated Si(111)/Cu(2, 4, and 7)/Pd(3)/Co(0.7)/Pd(3) samples. For brevity, a sample with a Cu buffer thickness of X nm is denoted Cu(X). The results are compared with previously investigated Si(111)/Cu(2)/Pd(3–15)/Co(0.7)/Pd(3) structures [19], denoted Pd(X), where X is the bottom Pd thickness. Below, the samples with variable Cu buffer thickness are referred to as the Cu-series, and those with variable bottom Pd thickness as the Pd-series.

The Cu buffer layer serves not only to induce interfacial roughness but also to suppress Pd silicide formation during growth and to promote (111)-oriented Pd epitaxy. A thin Cu silicide layer forms at the initial stages of Cu deposition on Si [22]. Starting at a nominal Cu thickness of approximately 0.6 nm, pure Cu(111) grows as flat two-dimensional islands in a layer-by-layer mode. Near a thickness of 3 nm, the growth changes to an island-growth mode, as indicated by the appearance and subsequent disappearance of oscillations in the intensity of the RHEED specular reflection recorded during Cu deposition on Si(111) (Fig. 1).

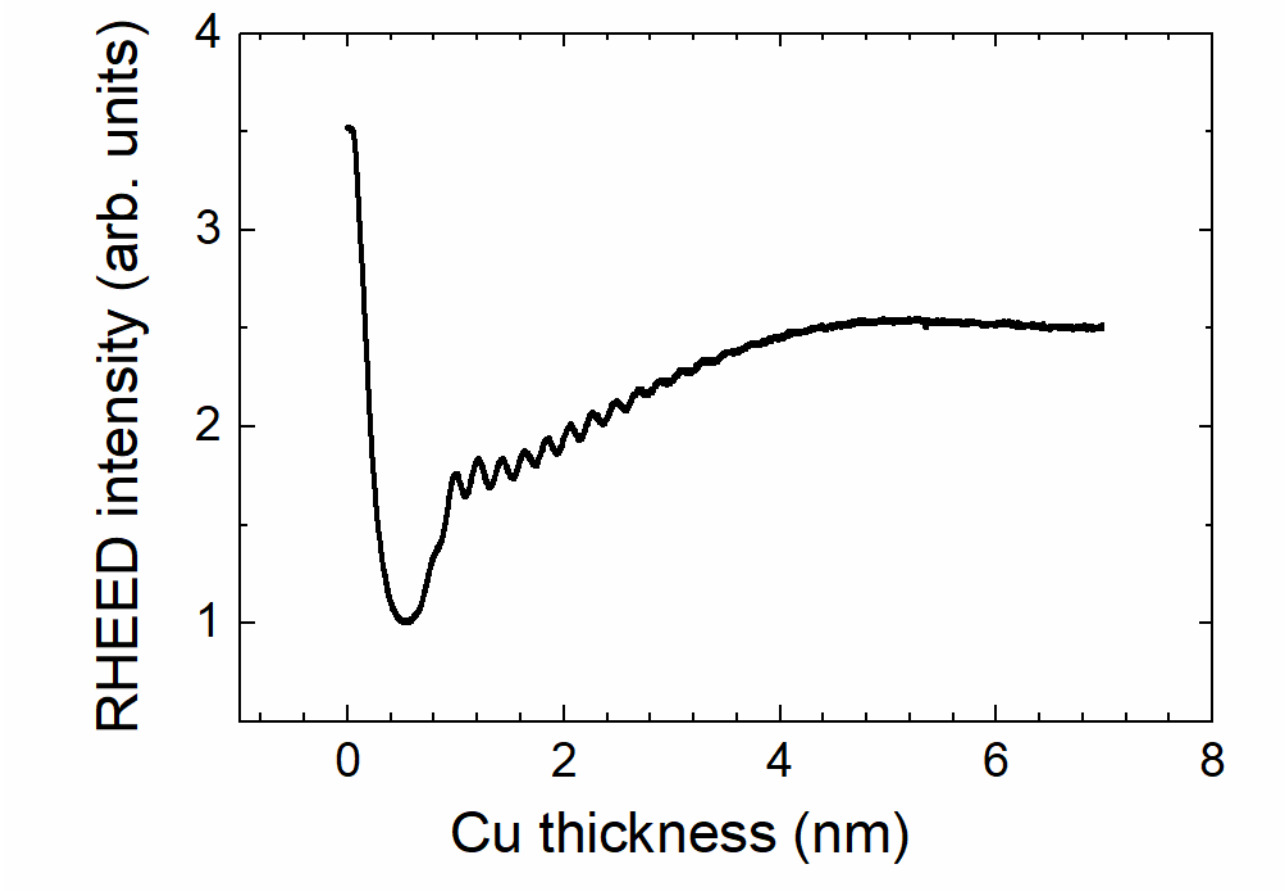

**Fig. 1** Oscillations of the RHEED specular-reflection intensity recorded during Cu deposition on Si(111)

The epitaxial orientation relationships determined from RHEED are Cu(111) ‖ Si(111) and Cu[11−2] ‖ Si[10−1]. These relationships are maintained during subsequent Pd and Co deposition. Figure 2 shows RHEED patterns from the Cu, bottom Pd, and Co surfaces in the Cu(2) and Cu(7) samples. The Cu lattice parameter already reaches its bulk value at a Cu thickness of 2 nm [22]. Consequently, relaxation of the lattice parameters of the subsequently deposited Pd and Co layers is not expected to depend on Cu buffer thickness.

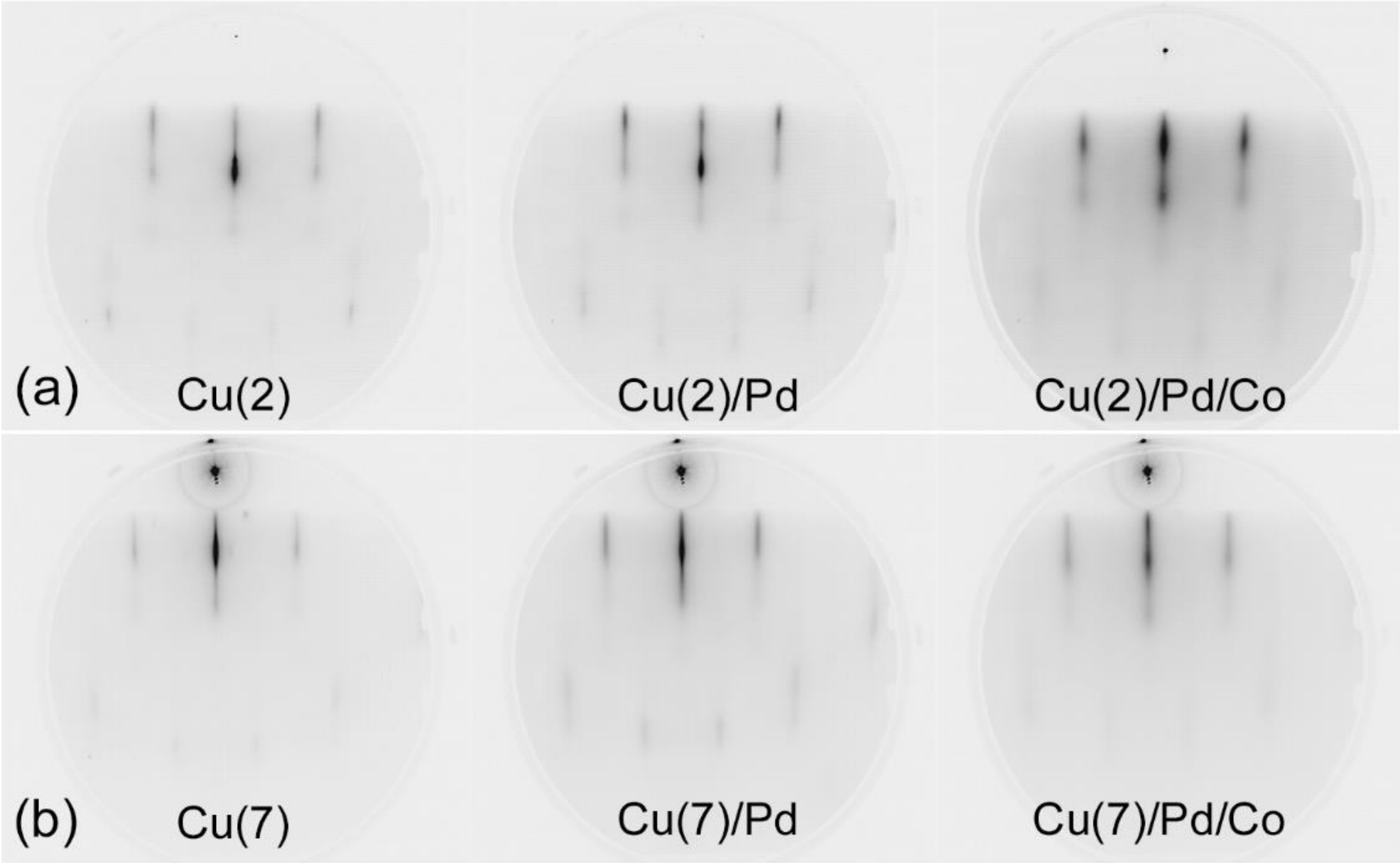

**Fig. 2** RHEED patterns from the surfaces of Cu, bottom Pd, and Co layers in (a) Cu(2) and (b) Cu(7) samples

Figure 3 shows STM images of the Cu, Pd, and Co surfaces in the Cu(2), Cu(4), and Cu(7) samples. Increasing the Cu thickness from 2 to 4 nm transforms the initially flat, two-dimensional islands into three-dimensional structures (Fig. 3(a-b)). The Cu islands in the Cu(4) sample have a hexagonal shape and are connected by narrow necks near their bases. With further Cu deposition, both the height and lateral dimensions of the islands increase (Fig. 3(c)). Pd tends to form islands with a shorter characteristic periodicity on the Cu surface (Fig. 3(d)). Pd atoms accumulate at the edges of the larger Cu islands, tending to stabilize their triangular shape. At the same time, smaller Pd islands form in the depressions between the Cu islands, as clearly visible in Figs. 3(e) and 3(f). As previously shown for the Pd-series [19], Co deposited on bottom Pd tends to occupy the spaces between Pd islands (Fig. 3(g)). The same behavior is observed in the Cu-series. In both the Cu(4) and Cu(7) samples, Co deposition increases the area occupied by the small islands and restores their hexagonal appearance (Figs. 3(h) and 3(i)). However, the nominal Co coverage is too small to substantially modify the pronounced surface topography generated by Cu and Pd.

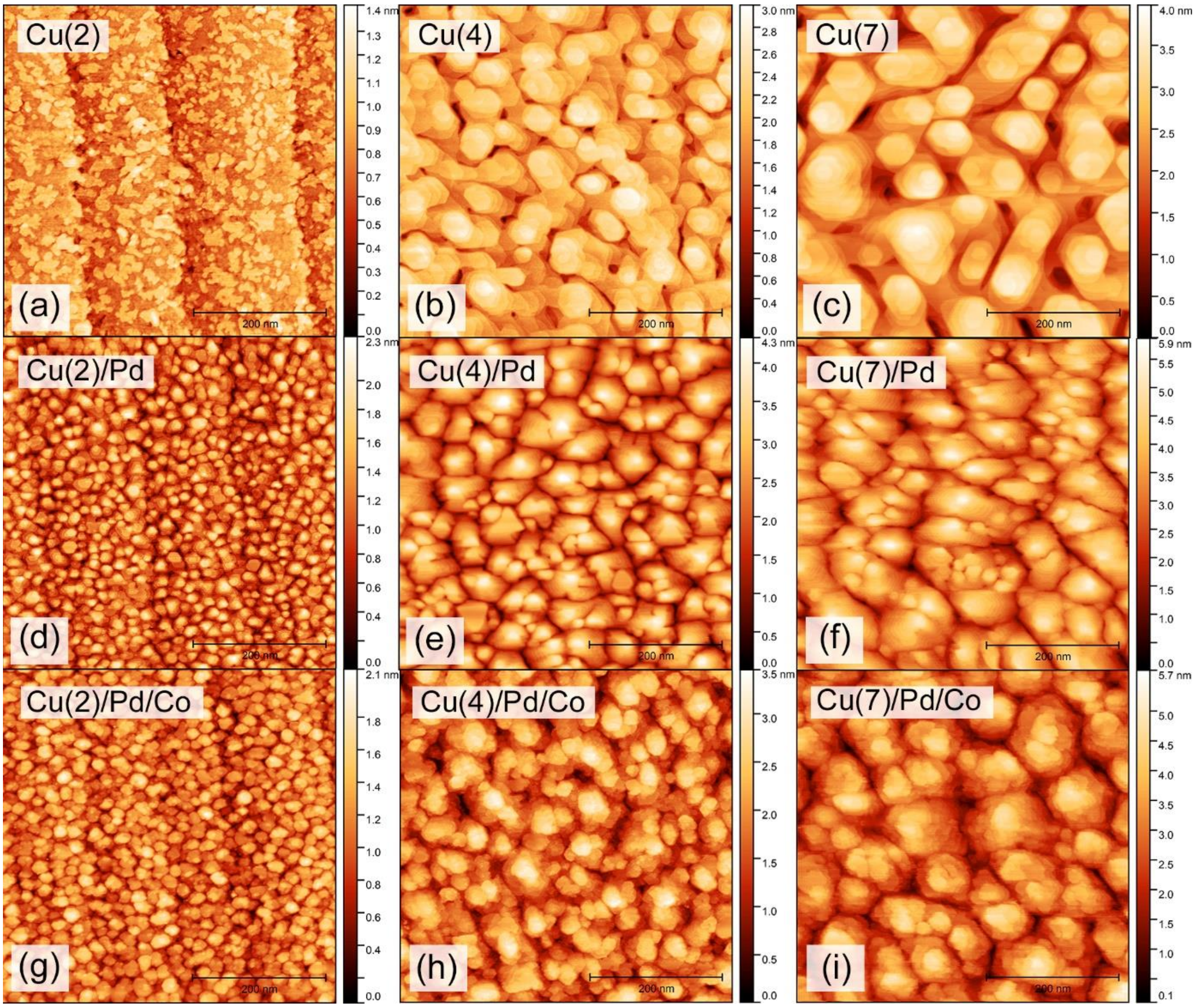

**Fig. 3** STM images of the (a)–(c) Cu, (d)–(f) bottom Pd, and (g)–(i) Co surfaces in the Cu-series samples. Each image covers an area of 500 × 500 nm²

Figure 4(a) compares the dependence of the rms surface roughness, $S_q$, on Cu thickness in the Cu-series and on bottom Pd thickness in the Pd-series. As the Cu layer thickness increases from 2 to 7 nm, the rms roughness of the Cu surface rises from 0.12 to 0.63 nm. Deposition of Pd on Cu significantly increases $S_q$: for the corresponding Cu thickness range, the roughness of the Pd surface increases from 0.25 to 0.84 nm. This behavior can be attributed to the sharpening of Cu islands by Pd deposited on their flat upper terraces, as well as the formation of additional small Pd islands. Due to its low nominal coverage, Co deposition does not substantially alter $S_q$. Although the rms roughness in the Cu-series is slightly larger than in the Pd-series, the values remain comparable. Notably, a similar roughness amplitude can be generated using a considerably thinner Cu buffer than the Pd layer required in the Pd-series.

The principal morphological distinction between the Cu- and Pd-series is the characteristic island periodicity. The average periodicity was obtained from the two-dimensional height autocorrelation function averaged over all in-plane directions. Figure 4(b) compares its dependence on Cu thickness with the corresponding results for the Pd-series [19]. In the Pd-series, the island periodicity increases from approximately 30 to 38 nm as the bottom Pd thickness increases from 3 to 15 nm; for comparison, we use the series-averaged value of 34 nm. In the Cu-series, the characteristic surface periodicity is primarily established during Cu growth and is only weakly modified by Pd deposition. Co partially fills the spaces between islands and increases the measured periodicity, particularly in the Cu(4) sample. Averaging the periodicity over the investigated interfaces gives approximately 70 and 90 nm for Cu(4) and Cu(7), respectively, corresponding to factors of 2.06 and 2.65 relative to the Pd-series average.

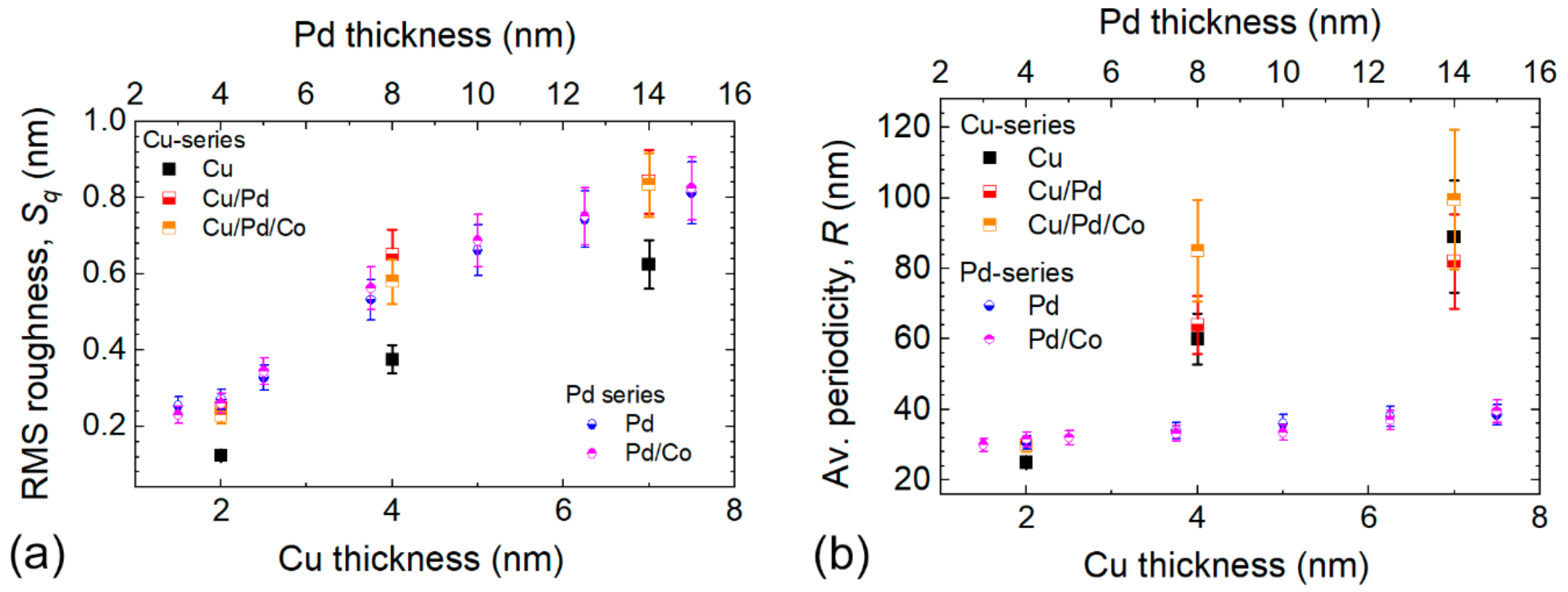

**Fig. 4** Dependence of (a) rms roughness and (b) average surface-feature periodicity on Cu thickness for the Cu-series and bottom Pd thickness for the Pd-series. Data for Pd-series was taken from Ref. [19] for comparative analysis

## 2.2 Magnetic Properties and Perpendicular Magnetic Anisotropy

Figures 5(a) and 5(b) show the magnetic hysteresis loops of the Cu(2), Cu(4), and Cu(7) samples measured with the field applied perpendicular and parallel to the film plane, respectively. The coercive field, $H_c$, increases markedly with Cu buffer thickness, whereas the effective PMA energy density, $K_{eff}$, decreases slightly. The dependencies of the coercive field and PMA field on Cu thickness are shown in Figs. 5(c) and 5(d), along with the corresponding results for the Pd-series. Both the effective PMA field, $H_a$, and coercive field in the Cu-series are comparable to those in the Pd-series. The saturation magnetization remains constant within experimental uncertainty, with $M_s$ = 1570 kA/m. The DW anisotropy field, $H_K$, was calculated following Ref. [23]:

$$H_K = M_s \frac{d_{Co}}{\pi\Delta} \ln 2 = M_s \frac{d_{Co}}{\pi} \sqrt{\frac{K_{eff}}{A}} \ln 2 ,$$

(1) where $d_{Co}$ is the Co thickness, $\Delta = \sqrt{\frac{A}{K_{eff}}}$ is the DW thickness parameter. The exchange constant of the epitaxial Co used to calculate the thickness of DWs was taken as 25 pJ/m [24]. The DW anisotropy field was ≈ 57 mT. The magnetic parameters of the samples including those derived by the analysis of the hysteresis loops are indicated in **Table 1**.

**Table 1** Magnetic parameters of the samples

| | $\mu_0 H_c$ (mT) | $\mu_0 H_a$ (T) | $M_s$ (MA/m) | $K_{eff}$ (MJ/m$^3$) | Δ (nm) | $\mu_0 H_K$ (mT) | $\mu_0 H_D$ (mT) |
|---|---|---|---|---|---|---|---|
| Cu(2) | 11 | 1.2 | 1.59 | 0.94 | 5.2 | 59 | 27 |
| Cu(4) | 13.5 | 1.06 | 1.57 | 0.83 | 5.5 | 56 | 30 |
| Cu(7) | 19 | 1.05 | 1.57 | 0.82 | 5.5 | 55 | 32 |

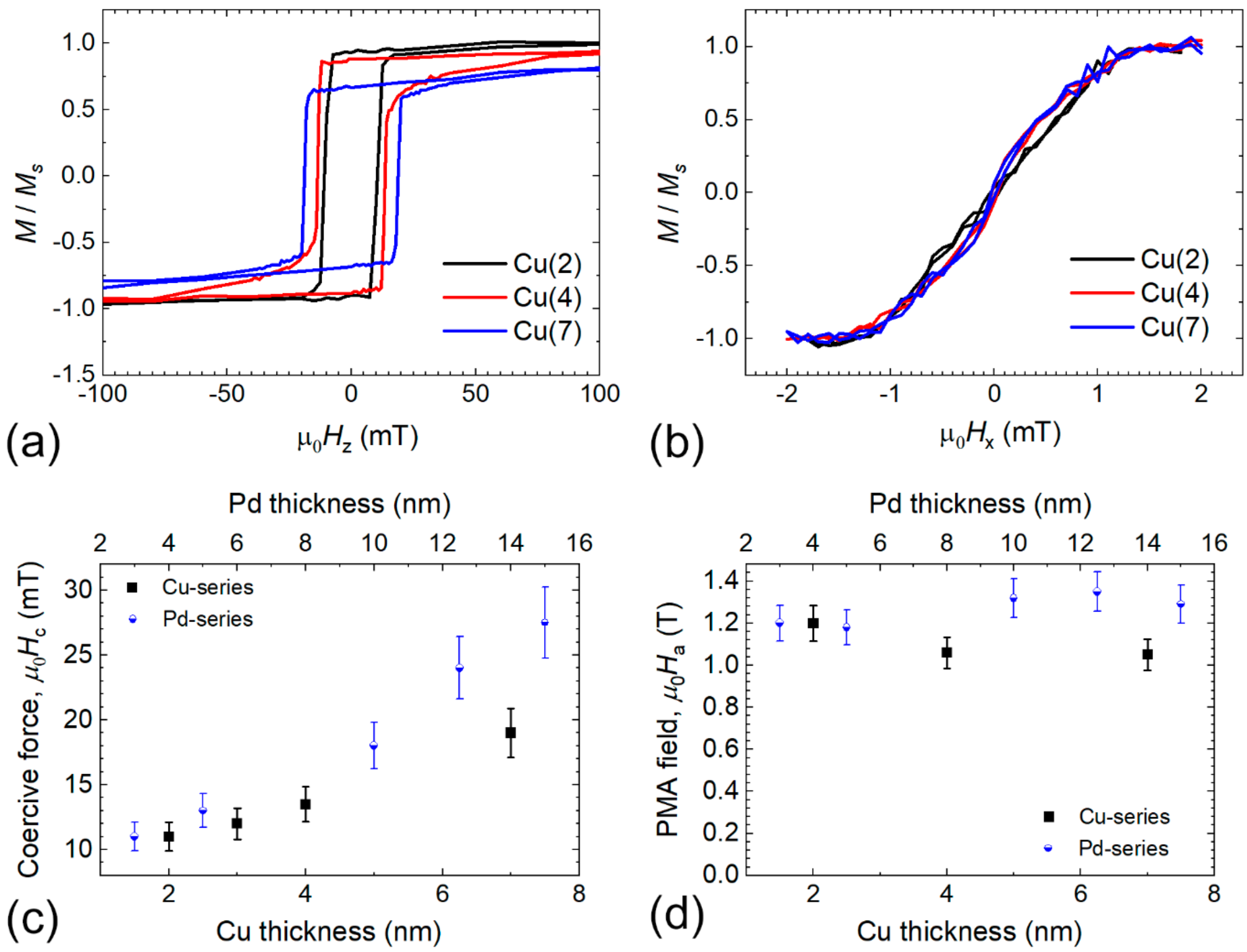

**Fig. 5** Magnetic hysteresis loops of the Cu-series measured with magnetic fields applied (a) perpendicular and (b) parallel to the film plane. Dependence of (c) coercive field and (d) effective PMA field on Cu thickness for the Cu-series and on bottom Pd thickness for the Pd-series. Data for Pd-series was taken from Ref. [19] for comparative analysis

## 2.3 Domain-Wall Dynamics

Figure 6(a) illustrates the expansion of magnetic domains in the Cu-series of samples under simultaneous OOP and IP magnetic-field pulses. The white differential magnetic contrast represents the newly reversed area, whereas the central gray region corresponds to the initial reversed domain, whose contrast was subtracted before analysis. The arrows indicate the magnetization directions in the nucleated domain and its surroundings. The driving OOP field was directed along the positive z axis (↑), and the IP field was 100 mT. Different OOP fields were used for the individual samples because the coercive field depends strongly on Cu thickness. As Cu thickness increases, the mobility of the ↑↓ DW increases, whereas that of the ↓↑ DW decreases.

The dependence of DW velocity on the IP magnetic field for the Cu-series is shown in Fig. 6(b). For ideal alignment of the IP electromagnet, the velocities of the opposite domain boundaries obey $v_{\uparrow\downarrow}(H_x) = v_{\downarrow\uparrow}(-H_x)$. To reduce random measurement errors, the corresponding velocities were averaged according to $v_{\uparrow\downarrow,av}(H_x) = v_{\downarrow\uparrow,av}(-H_x) = [v_{\uparrow\downarrow}(H_x) + v_{\downarrow\uparrow}(-H_x)]/2$. The driving OOP field was selected separately for each sample so that the measured velocity ranges were approximately comparable, while DW motion remained in the creep regime in every sample (See Supplementary Information). DW velocities were determined from the intersection of the expanded-domain boundary with the horizontal line connecting the leftmost and rightmost points of the initial circular domain. As shown in Fig. 6(b), the velocity of the ↑↓ DW increases with positive $H_x$ for all Cu thicknesses, with a change in slope near 50–70 mT. By contrast, the ↓↑ DW velocity decreases, reaches a minimum near 50–70 mT, and subsequently increases. This behavior is associated with a change in the DW magnetic configuration induced by the IP field. The DMI fields were estimated from $v(H_x)$ dependencies measured in the depinning regime, where the strong velocity asymmetry is suppressed and the positions of the velocity minima can be used to estimate the DMI field [25, 26, 20]. The DMI fields in the Cu-series were found to be small and were neglected in the subsequent analysis of the DW dynamics. The corresponding results are presented in Supplementary Information. Accordingly, the chirality of an ↑↓ DW changes from left-handed to right-handed as the IP field is varied over the interval from $-H_K$ to $+H_K$. The calculated anisotropy field is approximately 57 mT for all samples, consistent with the fields at which the kinks appear in the measured $v(H_x)$ curves.

The difference between the mobilities of DWs with opposite chirality increases with Cu thickness and the accompanying increase in interface roughness. This observation indicates that the roughness of the magnetic interfaces plays an important role in the giant mobility asymmetry, irrespective of whether it is generated by Cu or Pd growth. To quantify the mobility difference, we calculated the asymmetry coefficient $A(H_x) = v_{\uparrow\downarrow}(H_x)/v_{\downarrow\uparrow}(H_x)$ for positive IP fields (Fig. 6(c)). The $A(H_x)$ curves approach saturation for $H_x$ above approximately 90 mT. The asymmetry coefficients averaged over $H_x > 90$ mT, denoted $A_{sat}$, provide a quantitative measure of the mobility asymmetry. Figure 6(d) compares $A_{sat}$ for the Cu-series as a function of Cu thickness and the Pd-series as a function of bottom Pd thickness on a logarithmic scale. The enhancement of DW mobility asymmetry with increasing Cu buffer thickness is considerably stronger than that achieved by increasing the bottom Pd thickness. For Cu(7), we observed a velocity ratio of approximately 50,000 between right- and left-handed DWs, more than an order of magnitude larger than the highest reliably measured asymmetry in the Pd-series. We note, however, that accurate velocity measurements for the Pd(15) sample were not possible because expanding domains merged during the measurements.

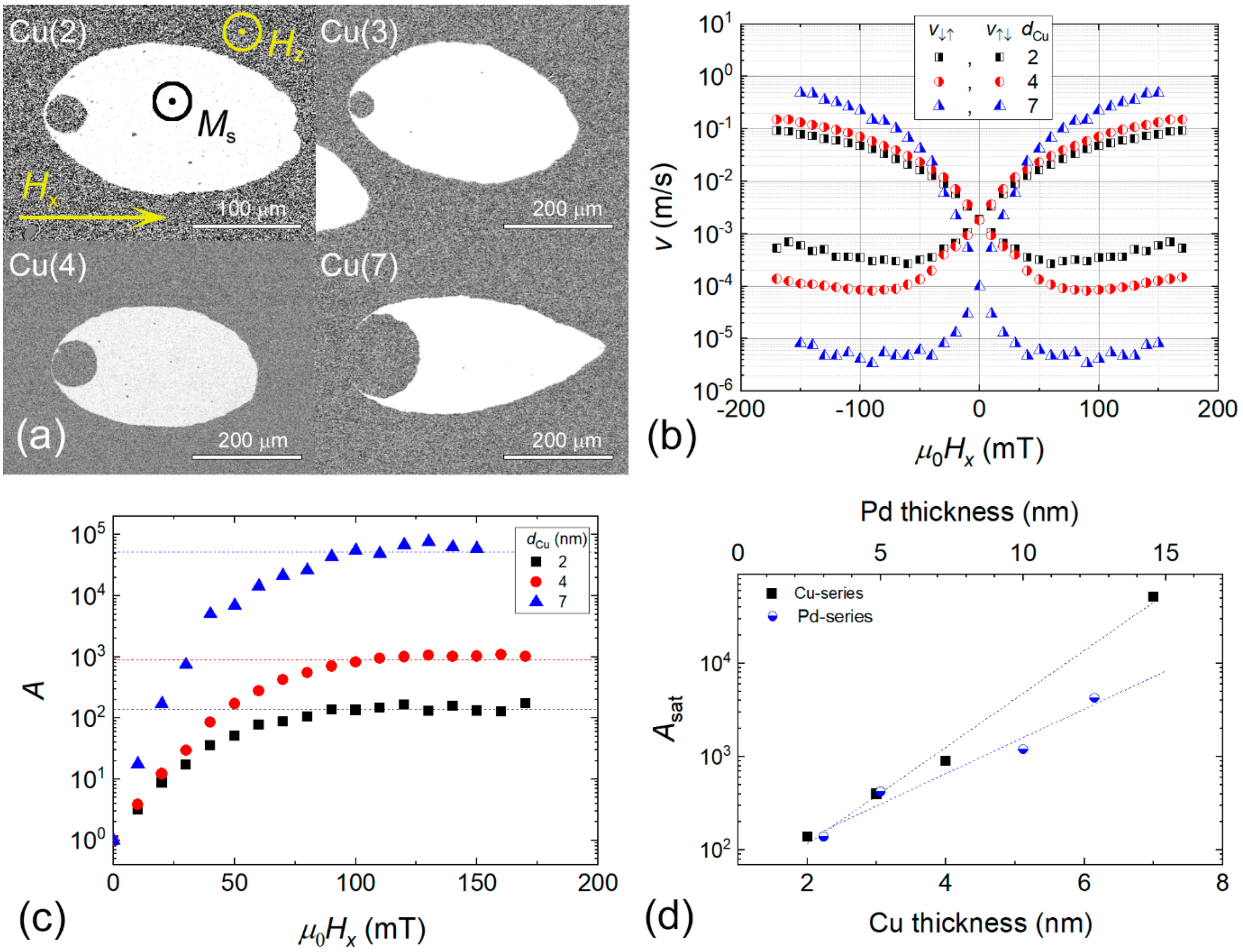

**Fig. 6** (a) Domain expansion in samples with different Cu buffer layer thicknesses following combined IP and OOP magnetic-field pulses of sample-dependent duration. White differential contrast denotes the area added to the initial domain. (b) Dependencies of the velocities of the left (↓↑) and right (↑↓) DWs on the IP magnetic field measured in the OOP magnetic fields of 14, 15, and 21 mT in the Cu(2), Cu(4), and Cu(7) samples, respectively. (c) Asymmetry coefficient $A$ as a function of the IP magnetic field. (d) Effective saturation asymmetry $A_{sat}$ versus Cu thickness, with the corresponding Pd-series data from Ref. [19]

## 3. Discussion

The results for the Cu- and Pd-series demonstrate that the asymmetry of DW mobility in epitaxial Pd/Co/Pd(111) heterostructures cannot be characterized by the rms roughness of the magnetic interfaces alone. Although the two series exhibit comparable rms roughness values and magnetic parameters, the Cu-series shows a considerably stronger enhancement of the mobility asymmetry, reaching a velocity ratio of approximately 50,000 in Cu(7). As shown in Fig. 6(d), the slope of log $A_{sat}$ versus $d_{Cu}$ is 3.7 times larger than that of log $A_{sat}$ versus $d_{Pd}$.

The main structural difference between the two series is the lateral organization of the interfacial morphology. In the Pd-series, increasing the bottom Pd thickness leads to the growth of Pd islands with a characteristic lateral periodicity of approximately 30–38 nm. By contrast, the morphology of the Cu-series is primarily determined by three-dimensional growth of the Cu buffer layer. Subsequent Pd deposition modifies this morphology by forming smaller islands on the Cu surface and within the trenches separating the larger Cu islands. Consequently, the Cu-series exhibits a hierarchical interfacial morphology comprising features with different lateral length scales. The characteristic periodicity reaches approximately 90 nm in the Cu(7) sample.

We propose that this difference in the spatial organization of interfacial roughness is responsible for the different efficiencies with which the DW mobility asymmetry can be controlled in the two series. The relatively large Cu-induced morphological features may generate spatially correlated variations in local Co thickness, PMA, and DW energy. Such variations can affect the collective interaction of DWs with pinning disorder, including the characteristic activation barriers and attempt frequencies governing thermally activated propagation.

Because the DW velocity in the creep regime depends exponentially on the activation barrier, even relatively small chirality-dependent changes in the pinning energy landscape can produce exceptionally large differences between the velocities of DWs with opposite chirality. Thus, the enhanced mobility asymmetry in the Cu-series may originate from more efficient coupling of the chiral magnetic structure of DWs to spatially correlated disorder induced by the Cu buffer layer.

Importantly, interfacial roughness does not itself select a preferred DW chirality. Rather, we suggest that the specific Cu-induced morphology amplifies an existing chirality-sensitive contribution to DW pinning. This interpretation is consistent with the pronounced mobility asymmetry observed in the creep regime and its suppression upon approaching the flow regime in previously studied Pd/Co/Pd(111) films [20].

Three-dimensional islands on a crystalline surface generally consist of terraces separated by monoatomic steps. If islands of comparable height and shape become wider, the average separation between monoatomic steps increases, reducing the areal density of elementary structural defects. On the Cu buffer layer surface, the characteristic terrace width, corresponding to the separation between monoatomic steps, is of the order of a few nanometers, whereas the island periodicity is tens of nanometers. It remains unclear which structural defects dominate DW pinning in the present system: monoatomic steps or regions of locally nonuniform Co thickness whose spacing reflects the periodicity of the large islands. In either case, we hypothesize that an increase in the characteristic spacing between pinning centers may favor a crossover from collective pinning to pinning at individual defects.

If DWs are pinned predominantly by monoatomic steps, the relevant quantity is the ratio of the terrace width to the DW width. Géhanne et al. showed that in systems with weak collective pinning, the effective disorder correlation length is determined by the DW width, over which many weak pinning sites are averaged [27]. In the present system, the characteristic DW width is approximately 6 nm. Increasing the spacing between steps could therefore reduce the number of defects averaged across the DW width and promote a crossover toward pinning on isolated defects. In the investigated samples, the creep-regime dependencies of ln $v$ are linear in $(\mu_0 H_z)^{-0.25}$, which is compatible with this interpretation [28]. However, observing a creep exponent of 0.25 alone is not sufficient to establish weak collective pinning.

Alternatively, if DWs are pinned by spatially separated regions of nonuniform Co thickness in the depressions between larger islands, a defect several nanometers across may interact with a DW over a length comparable to its width, resembling localized pinning. In that case, however, one must also consider how many such defects lie along the shortest approximately rigid DW segment, characterized by the Larkin length. Following Jeudy et al. [29], the Larkin length, which defines the characteristic length of an elastically correlated DW segment, can be estimated as

$$L_c = \sqrt{\frac{\sigma\xi}{2M_s\mu_0 H_d}}, \qquad (2)$$

where $\sigma$ is the DW energy per unit area, $\xi$ is the disorder correlation length, taken here to equal the defect periodicity, and $H_d$ is the depinning field. The depinning fields for the samples were determined from fitting of the $v(H_z)$ dependencies by methodic described in Refs. [29, 27] (See Supplementary Information). Taking $\sigma = 4(AK_{eff})^{1/2}$ and $\xi$ = 28, 70, and 90 nm for Cu(2), Cu(4), and Cu(7), respectively, we obtain $L_c$ = 81, 117, and 130 nm. As the interface periodicity increases, fewer pinning features are contained within a Larkin length, which may favor a crossover toward pinning dominated by individual defects.

Although the microscopic mechanisms responsible for DW pinning in this system remain to be established, the present findings suggest that engineering both the amplitude and lateral spatial organization of interfacial roughness provides an additional route to controlling chirality-dependent DW mobility in epitaxial magnetic heterostructures. Further measurements of creep exponents, nonuniversal creep parameters, and their dependence on the IP magnetic field are needed to distinguish the mechanisms by which the characteristic roughness periodicity influences chirality-sensitive DW pinning.

## 4. Conclusions

We have investigated the chirality-dependent creep dynamics of DWs in epitaxial Si(111)/Cu(2–7)/Pd(3)/Co(0.7)/Pd(3) films. The magnetic configuration and chirality of the DWs were varied by applying an IP magnetic field, while an OOP magnetic field drove their propagation. DWs of opposite chirality were found to exhibit markedly different mobilities. By changing the Cu buffer thickness, we controlled both the rms amplitude and characteristic lateral periodicity of the interfacial roughness. Comparison of Cu/Pd/Co/Pd samples with roughness generated by varying either the Cu buffer thickness or the bottom Pd thickness revealed an order-of-magnitude increase in the DW mobility asymmetry as the characteristic roughness periodicity increased from approximately 34 to 90 nm, while the other structural and magnetic parameters remained broadly comparable. These results demonstrate that the giant chirality dependence of DW mobility in Pd/Co/Pd can be efficiently tailored by engineering not only the amplitude of interfacial roughness but also its spatial periodicity.

**Funding**

The work of A.V. Davydenko, N.N. Chernousov, A.A. Turpak, and A.S. Pashenko was supported by the Russian Scientific Foundation (grant No. 25-22-20079, https://rscf.ru/project/25-22-20079) and the Ministry of professional education and employment of the population of Primorsky krai (grant no. 26-810-62470-2-0063-0000011).

**Supplementary Information for the manuscript**

**Estimation of the depinning and Dzyaloshinskii-Moriya interaction fields**

The dependence of the domain wall (DW) velocity on the out-of-plane (OOP) magnetic field was measured for each sample to establish the appropriate OOP field for subsequent in-plane (IP) field measurements in the creep regime (Fig. S1). The depinning field was determined by fitting the experimental data to the creep and depinning law, following the self-consistent analysis methodology originally proposed by Jeudy *et al.* [S1]. The experimental data were fitted using the creep and depinning laws:

$$v(H) = v(H_d)\exp\left(-\frac{\Delta E}{k_B T}\right), \qquad \text{Creep law: } H < H_d \qquad \text{(S1)}$$

$$v(H) = \frac{v(H_d)}{x_0}\left(\frac{T_d}{T}\right)^{\psi}\left(\frac{H-H_d}{H_d}\right)^{\beta}, \qquad \text{Depinning law: H } \gtrsim H_d \qquad \text{(S2)}$$

where $\Delta E = k_B T_d((H/H_d)^{-\mu} - 1)$ is the energy barrier of the creep regime and the depinning law is the asymptotic velocity behavior, $\mu = 1/4$, $\beta = 0.25$ and $\psi = 0.15$ are universal critical exponents and $x_0 = 0.65$ is a universal constant characterizing the quenched Edwards Wilkinson universality class [S2]. The nonuniversal parameters, which are fitting, are the characteristic height of effective pinning barrier $k_B T_d$ and coordinates of the depinning threshold $H_d$ and $v(H_d)$. The depinning fields $H_d$ extracting from these dependencies: 27 mT for Cu(2), 30 mT for Cu(4), and 32 mT for Cu(7).

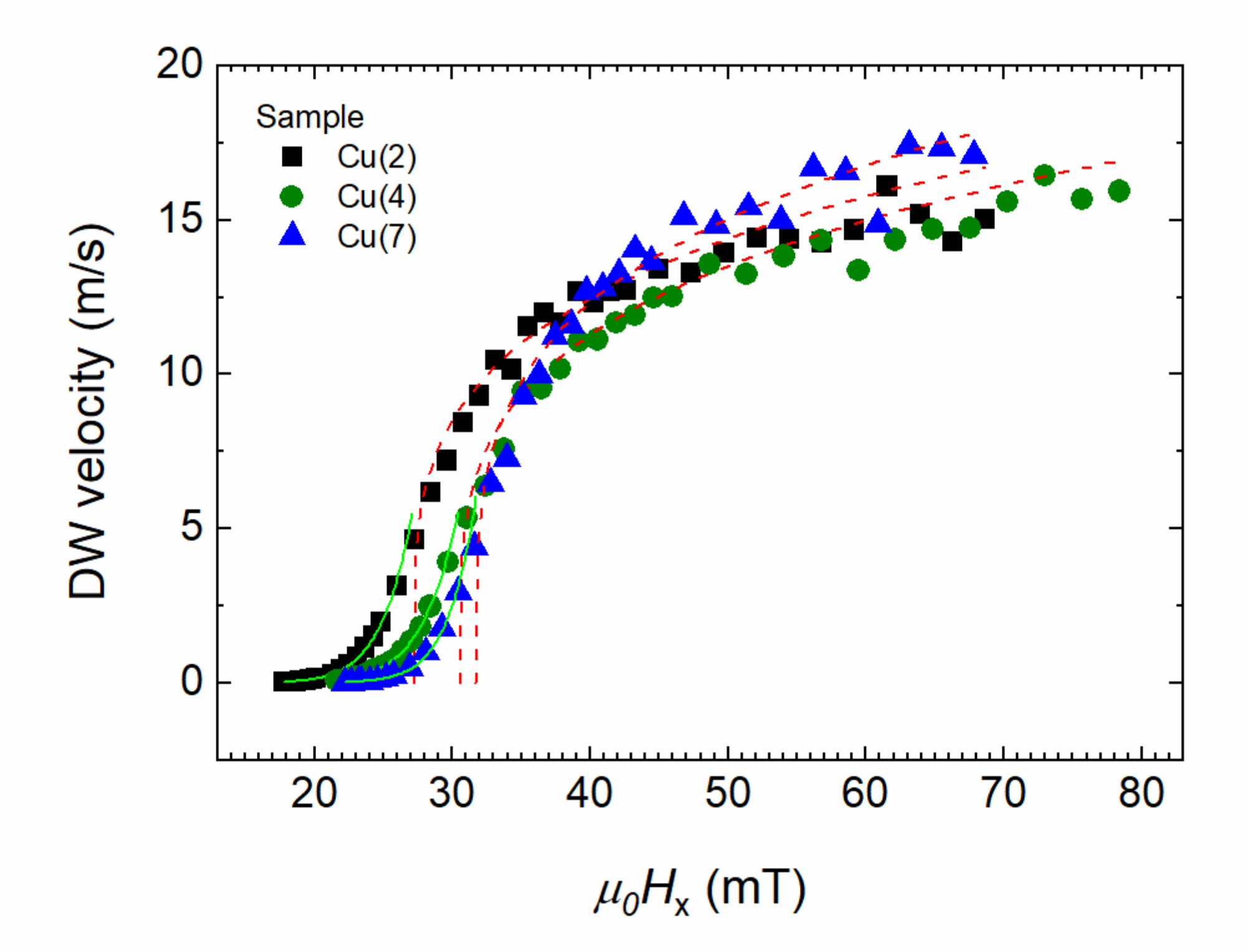

**Fig. S1** Velocity of DW as a function of the OOP field for Cu(2) (black squares), Cu(4) (green circles), and Cu(7) (blue triangles). The solid green and dashed red curves correspond to the fits of the creep and depinning laws.

The weak Dzyaloshinskii-Moriya interaction (DMI) in these samples was confirmed by measuring the DW velocity dependence on the IP field in the depinning regime under high OOP fields (Fig. S2). It is well established that, at high OOP fields in the depinning or flow regime, the asymmetry in the $v(H_x)$ dependencies disappears [S3]. Hence, the IP magnetic field corresponding to the minimum DW velocity equals the effective DMI field. All $v(H_x)$ dependencies were fitted with the following equation [S4]

$$v \approx \frac{\Delta\gamma_0\alpha}{1+\alpha^2}\mu_0 H_z\left(1+\left(\frac{\pi}{2}\right)^2\frac{(\mu_0 H_x+\mu_0 H_{\mathrm{DMI}})^2}{2(\mu_0 H_z)^2}\right)=$$

$$= c\mu_0 H_z\left(1+\left(\frac{\pi}{2}\right)^2\frac{(\mu_0 H_x+\mu_0 H_{DMI})^2}{2(\mu_0 H_z)^2}\right), \qquad \text{(S3)}$$

where $\Delta$ is a DW thickness, $\gamma_0$ is a gyromagnetic ratio of Co, $\alpha$ is a damping factor, c is a fitting parameter that does not depend on the IP magnetic field ($\Delta(H_x)$ is considered to be weak and neglected). The fitting yields the DMI field values: 2 mT for Cu(2), 5 mT for Cu(4), and 8 mT for Cu(7).

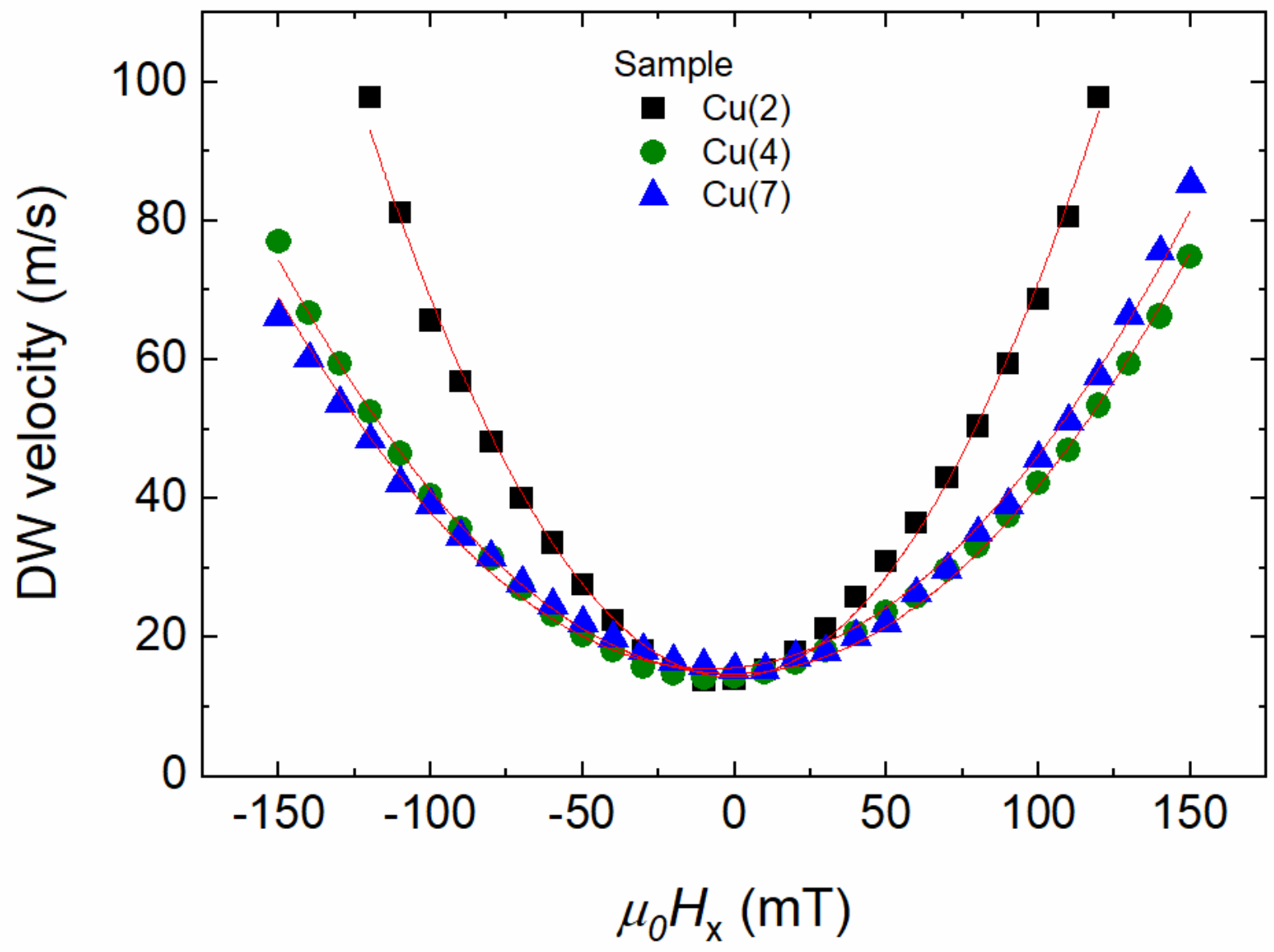

**Fig. S2** Velocity of ↓↑ DW as a function of the in-plane magnetic field for Cu(2) (black dots), Cu(4) (green dots), and Cu(7) (blue dots) measured in the OOP magnetic fields of 50, 61, and 52 mT, respectively